# O. CHERTOV, D. TAVROV

# PROVIDING DATA GROUP ANONYMITY USING CONCENTRATION DIFFERENCES

**Abstract.** *Public access to digital data can turn out to be a cause of undesirable information disclosure. That's why it is vital to somehow protect the data before publishing. There exist two main subclasses of such a task, namely, providing individual and group anonymity. In the paper, we introduce a novel method of protecting group data patterns. Also, we provide a comprehensive illustrative example.*

**Анотація.** *Вільний доступ до цифрових даних може призводити до небажаного витоку інформації. Саме тому потрібно деяким чином захищати дані перед оприлюдненням. Існує два підвиди цієї задачі, а саме, забезпечення індивідуальної та групової анонімности. У роботі ми пропонуємо новітній метод захисту групових властивостей даних. Також наводиться ілюстративний приклад.*
**Ключові слова:** *групова анонімність, статистиний контроль за розкриттям інформації, вейвлет-перетворення.*

**Аннотация.** *Свободный доступ к цифровым данным может вести к нежелательной утечке информации. Поэтому, следует неким образом защищать данные перед публикацией. Существует два подвида этой задачи, а именно, обеспечение индивидуальной и групповой анонимности. В работе мы предлагаем новый метод защиты групповых свойств данных. Также приводится пример-иллюстрация.*
**Ключевые слова:** *групповая анонимность, статистический контроль за раскрытием информации, вейвлет-преобразование.*

## 1. Introduction

The data anonymity is a subject to researches in different fields, among which privacy-preserving data mining [1], statistical disclosure control [2], distributed privacy, cryptography and adversarial collaboration [3] can be mentioned. Moreover, the number of papers on this topic hasn't been reduced in the recent years (for instance, see incomplete but very demonstrative bibliography in [4]). It is mainly due to enhancing the public access to various data for the researchers (or other involved people). They can be possibly interested in obtaining either the data about health, insurance, and other personal information, or the large samples of complete surveys (e.g., census) [5, 6]. On the other hand, Sweeney showed in her classical works [7, 8] that mere depersonalizing the dataset along with excluding the identifiers (which unambiguously violate respondent's anonymity) from it isn't enough for privacy-preserving. That's why there is a need in more advanced methods for providing data anonymity which take into account information about other respondents. In practice, to provide data anonymity, different systems are used, μ-Argus being one of the most demonstrative. It was developed during the SDC-, CASC-, ESSnet-projects, and it's totally freeware [9].

But, if to analyze existing data anonymity methods more thoroughly, it comes out that they actually protect individual privacy only. In other words, they belong to the class of individual anonymity methods. At the same time, the problem of protecting respondent group distribution is still open. Let us consider a typical situation: we cannot mask the information about regional distribution of military personnel in terms of individual anonymity. Instead, we might complete this task by redistributing particular respondents over different regions to achieve needed patterns. But, there isn't any feasible algorithm developed yet to aid in our task.

In general, we can divide all known data anonymity methods into two large subclasses, namely, randomization and group-based anonymization methods.

The essential idea of the randomization methods is to mask records' attribute values by adding some noise to the data [1, 10]. In the situation described above, it is obvious that the added noise can certainly mask the true number of military personnel in a region. But, the distribution pattern (e.g., extreme numbers locations) will persist, because the noise should by default have a lot smaller amplitude than the signal itself.

On the other hand, group-based anonymization methods [11] aim mainly at gaining *k*-anonymity (using suppression, generalization, data swapping and so on). *K*-anonymity means that every attribute values' combination corresponds to at least *k* respondents in the dataset. In the case with our "military" example, we might mask individual information about, say, senior military officers (so that they cannot be distinguished among the others). But, since the key property ("military"/"civilian") is the only one available, splitting the population into these two groups with the follow-up anonymization within each of them doesn't lead to masking needed regional distribution.

Thus, we come to conclusion that the only acceptable option is indeed to (virtually) redeploy respondents between different regions. But, we also have to minimize possible loss of the resultant data utility.

## 2. The Aim of the Paper

In the paper, we discuss different ways of solving the group anonymity problem.

Moreover, we set a task even more complicated than the one described above. We consider a problem when the comparative distribution of two respondent groups' quantities (or ratios) is supposed to be protected. For that matter, we take young males and females distributed by regions and try to hide possible extreme differences between their ratios in each region. The reason for protecting this distribution is that such extremums can possibly reveal the location of some concealed military cantonment which isn't supposed to be known.

We propose to accomplish such a task by using wavelet transform (WT). It allows us to achieve needed patterns by redistributing wavelet approximation values. At the same time, fixing wavelet details and other features such as data mean value can surely prevent significant utility loss. To illustrate that, let us refer to [12]. In Russia, responses to 44 public opinion polls (1994-2001) yielded the following results. It turned out that the wavelet details actually reflect hidden time series features which can come in handy for sociological forecasting. And, last but not least, WT has been already used for providing data anonymity, though individual only [13] so far.

## 3. Theoretic Background

### 3.1. Group Anonymity Basics

Let's collect the depersonalized primary data into a so-called microfile (see Table 1).

Table 1. Microfile Data.

|          | $w_1$       | $w_2$       | $\ldots$ | $w_\eta$       |
|----------|-------------|-------------|----------|----------------|
| $r_1$    | $z_{11}$    | $z_{12}$    | $\ldots$ | $z_{1\eta}$    |
| $r_2$    | $z_{21}$    | $z_{22}$    | $\ldots$ | $z_{2\eta}$    |
| $\ldots$ | $\ldots$    | $\ldots$    | $\ldots$ | $\ldots$       |
| $r_\mu$  | $z_{\mu 1}$ | $z_{\mu 2}$ | $\ldots$ | $z_{\mu\eta}$  |

Here, μ stands for the number of respondents, η stands for the number of attributes; $w_j$ stands for the $j^{th}$ attribute, $r_i$ stands for the $i^{th}$ record, $z_{ij}$ stands for a microfile data element.

To protect important data patterns, we need to somehow redistribute particular elements $z_{ij}$. Let us formally define this task.

First of all, we need to distinguish which microfile elements we'll be eager to redistribute. Let's denote by $S_v$ a subset of a Cartesian product $w_{v_1} \times w_{v_2} \times ... \times w_{v_l}$ of Table 1 columns, where $v_i$, $i = \overline{1,l}$ are integers. This set will be called *a vital set*. Each vector from this set will be called *a vital value combination*. Respectively, we will call each element of such a vector *a vital value*, and $w_{v_i}$, $i = \overline{1,l}$ will be called *a vital attribute*.

We call these values such way because it is vital indeed to protect their distribution. In other words, attributes should be chosen as vital ones when the task is set to protect their distribution. E.g., if we wanted to hide the distribution of "Middle-aged women" we would need to take "Age" and "Sex" as vital attributes.

But, we may hide the "Middle-aged women" distributions over different value ranges. For instance, we can change their distribution over country regions, over ethnic groups, or even over the places they work at. Thus, let's denote by $S_p$ a subset of microfile data elements $z_{ip}$ corresponding to the $p^{th}$ attribute, $p \neq v_i \ \forall i = \overline{1, l}$. These elements will be called *parameter values*, whereas $p^{th}$ attribute will be called *a parameter attribute*. This attribute actually stands for a specific value range to redistribute vital values over.

In the case with "Middle-aged women", the parameter attribute could possibly be "Country region", "Ethnic group", or "Place of work".

Thus, providing group anonymity actually means redistributing records with vital value combinations over different parameter values.

After having defined the attributes mentioned above, we need to calculate the quantities of microfile records with every possible pair of a vital value combination and a parameter value. Received quantities can be gathered in an array of discrete values $q = (q_1, q_2, ..., q_m)$ which we will call *a quantity signal*.

As it was mentioned earlier, providing group anonymity has to be accomplished such way that data utility isn't reduced much. It can be easily achieved by using wavelet transform. If to modify wavelet approximation, but leave all the wavelet details either fixed or altered proportionally, we might fulfill the stated requirements.

Having applied these transformations to signal $q$, we receive a new quantity signal $\tilde{q} = (\tilde{q}_1, \tilde{q}_2, ..., \tilde{q}_m)$.

But, in many cases redistributing absolute quantities doesn't yield adequate results. Moreover, redistributing them may lead to a serious loss of data utility. Thus, modifying ratios sounds like a much better idea. That is why we need to modify our quantity signal by dividing its every value by the overall number of records with the same parameter value, but the vital values defining the superset for the records to be redistributed. For example, when redistributing "Middle-aged women" over the "Country regions", we might divide the middle-aged women quantities by the overall number of women in each region.

In the outcome we will receive *a concentration signal* $c = (c_1, c_2, ..., c_m)$. Then, performing operations identical to the case described above, we can get a new concentration signal $\tilde{c} = (\tilde{c}_1, \tilde{c}_2, ..., \tilde{c}_m)$ with a different distribution.

To show a bit more in detail how to modify a wavelet approximation, we need to revise WT basics first.

### 3.2. Necessary Wavelet Transform Basics

In this subsection we will revise only those wavelet theory facts which are required for the better understanding of our further explanations. You may find much more detailed information in [14, 15].

So, let's call an array of discrete values $s = (s_1, s_2, ..., s_m)$ *a signal*. Also, let a high-pass wavelet filter be denoted as $h = (h_1, h_2, ..., h_n)$, and a low-pass wavelet filter be denoted as $l = (l_1, l_2, ..., l_n)$.

Then, if we denote a convolution by $*$, and a dyadic downsampling by $\downarrow 2n$, we can perform signal $s$ one-level wavelet decomposition as follows:

$$a_1 = s *_{\downarrow 2n} l; \ d_1 = s *_{\downarrow 2n} h. \tag{1}$$

In (1), $s$ and $l$ (as well as $s$ and $h$) are being convoluted first, and then the result is being dyadically downsampled. In this case, $a_1$ is an array of approximation coefficients at level 1, whereas $d_1$ is an array of detail coefficients at level 1.

We can also apply (1) to $a_1$ and receive approximation and detail coefficients at level 2. Generally speaking, applying (1) to approximation coefficients at any level $k$-1 results in approximation and detail coefficients at level $k$:

$$a_k = a_{k-1} *_{\downarrow 2n} l = (\underbrace{(s *_{\downarrow 2n} l) ... *_{\downarrow 2n} l}_{k \ times}); \tag{2}$$

$$d_k = a_{k-1} *_{\downarrow 2n} h = (((s *_{\downarrow 2n} l) \dots *_{\downarrow 2n} l) *_{\downarrow 2n} h) \ . \qquad (3)$$

$$\underbrace{\phantom{(((s *_{\downarrow 2n} l) \dots *_{\downarrow 2n} l)}}_{k-1 \ times}$$

Every signal $s$ can be presented as a sum of approximation and details at appropriate levels:

$$s = A_k + \sum_{i=1}^{k} D_i \ . \qquad (4)$$

In (4), $A_k$ stands for an approximation at level $k$, and each $D_i$ stands for a detail at a particular level $i$. They are connected with the corresponding coefficients as follows:

$$A_k = ((a_k \underbrace{*_{\uparrow 2n} l) \dots *_{\uparrow 2n}}_{k \ times} l); \ D_k = (((d_k *_{\uparrow 2n} h) \underbrace{*_{\uparrow 2n} l) \dots *_{\uparrow 2n}}_{k-1 \ times} l) \ . \qquad (5)$$

In (5), $a_k$ and $d_k$ are being dyadically upsampled (which is denoted by $\uparrow 2n$) first, and then convoluted with an appropriate wavelet filter.

### 3.3. Wavelet Reconstruction Matrix

It may sound weird, but we cannot change wavelet approximation $A_k$ absolutely arbitrarily. This is mainly because the wavelet decomposition of a new signal $\tilde{s}$ (which is obtained as a sum of a new approximation and old details) in this case results in completely different details and approximation. Therefore, not a single detail is preserved.

The only opportunity to preserve the details is to alter approximation coefficients. According to (5), changing them doesn't influence the details at all.

But, formula (5) doesn't really suggest what coefficients should we change and how to receive a specific approximation. Fortunately, there exists another technique for obtaining approximations from coefficients.

In [15], it is described how to carry out WT using matrix multiplications only. In particular, we can present obtaining $A_k$ as follows:

$$A_k = M_{rec} \cdot a_k \ . \qquad (6)$$

We will call $M_{rec}$ a wavelet reconstruction matrix (WRM). We can always obtain it by consequent multiplications of appropriate upsampling and convolution matrices introduced in [15].

With the help of WRM, it is easy to find out what coefficients to change in order to get a specific approximation (an illustrative example is shown in the next section). After having defined new coefficients $\tilde{a}_k$, we can construct a new approximation (using either (5) or (6)). Then, we need to add this approximation to the initial wavelet details. As a result, we receive a new signal $\tilde{s}$ which totally suits our requirements.

### 3.4. Applying Concentration Differences to Obtaining New Data

There exist some real-life problems that cannot be solved by modifying a concentration signal corresponding to only one set of vital attributes. In these cases, the differences between different quantities (or ratios) are a subject to protection.

For that matter, we have to slightly extend our problem definition. Instead of defining one vital set we will define two such sets. The first one will be called *a main vital set*, and the other one will be called *a subordinate vital set*. We will call every vector from the main vital set *a main vital value combination*, and every element of this vector will be called *a main vital value*.

Respectively, every vector from the subordinate vital set will be called *a subordinate vital value combination*, whereas its every element will be called *a subordinate vital value.*

It is important to note that the parameter attribute remains common for both vital sets.

We can construct appropriate quantity and concentration signals. But, in this particular case, we won't even try to redistribute concentration signal values. We will construct an additional signal instead.

So, let $c^1 = (c_1^1, c_2^1, \dots, c_m^1)$ be *a main concentration signal* (built-up using main vital value combinations), and also let $c^2 = (c_1^2, c_2^2, \dots, c_m^2)$ be a corresponding *subordinate concentration signal*. Let us create *a concentration difference signal* as $\delta = (\delta_1, \delta_2, \dots, \delta_m) \equiv (c_1^1 - c_1^2, c_2^1 - c_2^2, \dots, c_m^1 - c_m^2)$ .

Our next step is to receive a new concentration difference signal $\tilde{\delta}$ . Afterwards, we can construct new concentration signals $\tilde{c}^1$ and $\tilde{c}^2$ which meet the following conditions:

1. The differences between these signals' values are $\tilde{\delta}$ elements.

2. New ratios don't differ from the initial ones significantly (for instance, the main concentration signal can stay fixed).

Using new concentration signals, we can always restore corresponding quantity signals. But, the mean values of these new signals will totally differ from the initial ones. This is totally unacceptable, because we cannot alter the overall number of records with appropriate vital values. To overcome this problem, we need to multiply the resultant quantity signals by such coefficients that guarantee preservation of the mean values. Due to the algebraic properties of convolution, in this case wavelet details of will be changed proportionally. And that completely satisfies our problem definition.

In the next section we will present a comprehensive example that will aid in better understanding the main steps of the described algorithm.

## 4. Experimental Results

We took Italy Census-2001 microfile provided by [5] as the data to analyze. This microfile contains various information on about 3 million respondents. To show the concentration differences method in action, we decided to set a suitable group anonymity task.

It is obvious that the differences between young males' and females' ratios can possibly point out the location of the Forze Armate Italiane cantonments. So, to mask these locations, we decided to choose the following parameter and value attributes.

We took "REGNIT" (which stands for "Region of Italy") as a parameter attribute because we aim at changing regional distribution of the mentioned ratios. Each attribute value stands for a particular region of Italy, except for the "1" value which stands for two regions, i.e. "Piedmont" and "Aosta Valley". For our purpose, we decided to split the data corresponding to this attribute proportionally using the official information about these two regions' population [16]. Further on, we will refer to "Piedmont" as "1P", and to "Aosta Valley" as "1V".

Eventually, we receive 20 parameter values standing for each region of Italy.

Since our task is to process the data corresponding to young males and females, we took "SEX" and "AGE" as both main and subordinate vital attributes. In the microfile we analyzed, age is grouped into categories, that's why we could take only one vital value that corresponds to the young age, i.e. "22". This value will serve as both main and subordinate one because we will redistribute males and females of the same age.

At the same time, we took "SEX" value "1" (standing for "Male") as a main value, whereas "2" ("Female") was chosen as a subordinate one.

Having determined the data to work with, we need to build up main and subordinate concentration signals. To perform that, we have to divide the number of young males and females in each region (see Table 2, the 3$^{rd}$ and the 5$^{th}$ rows) by the overall number of people living in the same region (see Table 2, the 2$^{nd}$ row). The resultant concentration signals are presented in the 4$^{th}$ and the 6$^{th}$ rows of Table 2.

Now, we can easily construct a concentration difference signal: $\delta = (0.0012, 0.0013, 0.0010, 0.0005, 0.0006, 0.0019, -0.0002, 0.0005, 0.0012, 0.0020, -0.0001, 0.0010, 0.0008, -0.0005, 0.0006, 0.0018, 0.0030, 0.0003, 0.0006, 0.0014)$.

In this paper, we present all the calculated numeric data with 4 decimal numbers (because of the limited space), though all the calculations were carried out with a higher proximity.

As we can see, there is a global maximum in the 17$^{th}$ signal value. Since this maximum can possibly expose the location of some military cantonment, we need to change the signal $\delta$ distribution. It can be accomplished using different approaches. For instance, we could transit the mentioned maximum to another region, or create other alleged maximums in different signal elements etc. For we would like to study how choosing different wavelet bases can help in choosing particular approach, we picked two wavelet bases to apply to our example, namely, the first and the second order Daubechies wavelet bases [14].

So, let us use the first order Daubechies low-pass wavelet filter $l^1 \equiv \left( \dfrac{1}{\sqrt{2}}, \dfrac{1}{\sqrt{2}} \right)$ to perform two-level wavelet decomposition (2) of a corresponding concentration difference signal: $a_2^1 = (0.0020, 0.0013, 0.0020, 0.0014, 0.0026)$.

According to (6), and using a suitable WRM (see Fig. 1a), we can obtain a signal approximation: $A_2^1 = (0.0010, 0.0010, 0.0010, 0.0010, 0.0007, 0.0007, 0.0007, 0.0007, 0.0010, 0.0010, 0.0010, 0.0010, 0.0007, 0.0007, 0.0007, 0.0007, 0.0013, 0.0013, 0.0013, 0.0013)$.

$$
\mathbf{M}_{rec}^{1} =
\begin{pmatrix}
0.5 & 0 & 0 & 0 & 0 \\
0.5 & 0 & 0 & 0 & 0 \\
0.5 & 0 & 0 & 0 & 0 \\
0.5 & 0 & 0 & 0 & 0 \\
0 & 0.5 & 0 & 0 & 0 \\
0 & 0.5 & 0 & 0 & 0 \\
0 & 0.5 & 0 & 0 & 0 \\
0 & 0.5 & 0 & 0 & 0 \\
0 & 0 & 0.5 & 0 & 0 \\
0 & 0 & 0.5 & 0 & 0 \\
0 & 0 & 0.5 & 0 & 0 \\
0 & 0 & 0.5 & 0 & 0 \\
0 & 0 & 0 & 0.5 & 0 \\
0 & 0 & 0 & 0.5 & 0 \\
0 & 0 & 0 & 0.5 & 0 \\
0 & 0 & 0 & 0.5 & 0 \\
0 & 0 & 0 & 0 & 0.5 \\
0 & 0 & 0 & 0 & 0.5 \\
0 & 0 & 0 & 0 & 0.5 \\
0 & 0 & 0 & 0 & 0.5
\end{pmatrix}
\qquad
\mathbf{M}_{rec}^{2} =
\begin{pmatrix}
0.6373 & 0 & 0 & 0 & -0.1373 \\
0.2958 & 0.2333 & 0 & 0 & -0.0290 \\
0.0792 & 0.4040 & 0 & 0 & 0.0167 \\
-0.0123 & 0.5123 & 0 & 0 & 0 \\
-0.1373 & 0.6373 & 0 & 0 & 0 \\
-0.0290 & 0.2958 & 0.2333 & 0 & 0 \\
0.0167 & 0.0792 & 0.4040 & 0 & 0 \\
0 & -0.0123 & 0.5123 & 0 & 0 \\
0 & -0.1373 & 0.6373 & 0 & 0 \\
0 & -0.0290 & 0.2958 & 0.2333 & 0 \\
0 & 0.0167 & 0.0792 & 0.4040 & 0 \\
0 & 0 & -0.0123 & 0.5123 & 0 \\
0 & 0 & -0.1373 & 0.6373 & 0 \\
0 & 0 & -0.0290 & 0.2958 & 0.2333 \\
0 & 0 & 0.0167 & 0.0792 & 0.4040 \\
0 & 0 & 0 & -0.0123 & 0.5123 \\
0 & 0 & 0 & -0.1373 & 0.6373 \\
0.2333 & 0 & 0 & -0.0290 & 0.2958 \\
0.4040 & 0 & 0 & 0.0167 & 0.0792 \\
0.5123 & 0 & 0 & 0 & -0.0123
\end{pmatrix}
$$

Fig. 1. Wavelet reconstruction matrices: a) WRM obtained using the first order Daubechies filter; b) WRM obtained using the second order one.

Also, we can obtain the details at levels 1 and 2 (using (5)), and sum them up: $D_1^1 + D_2^1 = (0.0002, 0.0003, 0.0000, -0.0005, -0.0001, 0.0012, -0.0009, -0.0002, 0.0002, 0.0010, -0.0012, 0.0000, 0.0002, -0.0012, -0.0001, 0.0011, 0.0016, -0.0010, -0.0007, 0.0001).$

As it follows from $M_{rec}^{1}$ (see Fig. 1a), changing one approximation coefficient results in altering 4 neighboring approximation values, and no other coefficient influences them. This means we can put some alleged maximums in our signal to solve the task, for we won't be able to eliminate signal's maximum totally.

In general, we can take any possible approximation coefficients, but for this particular example we decided to choose the following ones: $\tilde{a}_2^1 = (0.0036, 0.0018, 0.0019, 0.0018, 0.0009)$. Using these coefficients guarantees that the new signal's $17^{th}$ value will be lower than the present one, whereas the $6^{th}$ and the $10^{th}$ values will become similar to the $17^{th}$ one. This is exactly what we intended to do, i.e. not eliminate the initial maximum but create several other alleged ones.

Using (6), we can get a new approximation: $\tilde{A}_2^1 = (0.0016, 0.0016, 0.0016, 0.0016, 0.0009, 0.0009, 0.0009, 0.0009, 0.0010, 0.0010, 0.0010, 0.0010, 0.0009, 0.0009, 0.0009, 0.0009, 0.0004, 0.0004, 0.0004, 0.0004).$

By adding old details to a new approximation we can get a new concentration difference signal: $\tilde{\delta}^1 = (0.0018, 0.0018, 0.0016, 0.0011, 0.0008, 0.0021, -0.0000, 0.0007, 0.0011, 0.0019, -0.0002, 0.0010, 0.0010, -0.0003, 0.0008, 0.0020, 0.0021, -0.0005, -0.0003, 0.0005).$

As we see, we actually reached what we intended to. The next step is to construct new main and subordinate concentration signals that suit the requirements stated in the previous subsection. It can always be completed by solving a corresponding linear equation system with $2m$ unknowns and $m$ equations (these equations are the definitions of the $\delta$ elements).

We received the following ratios (of course, other solutions also are possible): $\tilde{c}_{(1)}^1 = (0.0269, 0.0268, 0.0283, 0.0304, 0.0282, 0.0269, 0.0210, 0.0251, 0.0265, 0.0290, 0.0285, 0.0296, 0.0318, 0.0319, 0.0369, 0.0381, 0.0343, 0.0363, 0.0349, 0.0339);$ $\tilde{c}_{(1)}^2 = (0.0251, 0.0250, 0.0267, 0.0293, 0.0274, 0.0249, 0.0211, 0.0244, 0.0254, 0.0271, 0.0287, 0.0287, 0.0308, 0.0322, 0.0361, 0.0361, 0.0323, 0.0368, 0.0352, 0.0334).$

Using these ratios and the quantities from the $2^{nd}$ row of Table 2, we can obtain new quantity signals $\hat{q}^1_{(1)}$ and $\hat{q}^2_{(1)}$.

But, the overall number of young males and females has been totally changed! To cope with this backfire, we need to multiply the quantity signals by appropriate coefficients, i.e. $\sum_{i=1}^{20} q^1_i / \sum_{i=1}^{20} \hat{q}^1_{(1)_i} = 0.9945$ and $\sum_{i=1}^{20} q^2_i / \sum_{i=1}^{20} \hat{q}^2_{(1)_i} = 0.9965$. The rounded results and the ratios calculated using the revised quantities are presented in Table 2 (rows 7 to 10).

After having solved the task using the first order wavelet filter, we propose to apply the second order one to see whether any other possibilities can show out.

So, let's take the second order Daubechies low–pass wavelet filter $l^2 \equiv \left( \dfrac{1+\sqrt{3}}{4\sqrt{2}}, \dfrac{3+\sqrt{3}}{4\sqrt{2}}, \dfrac{3-\sqrt{3}}{4\sqrt{2}}, \dfrac{1-\sqrt{3}}{4\sqrt{2}} \right)$ to perform two–level wavelet decomposition (2) of a concentration difference signal: $a^2_2 = (0.0021, 0.0017, 0.0018, 0.0010, 0.0029)$.

Using correspoding WRM from Fig. 1b, we get the following approximation: $A^2_2 = (0.0010, 0.0009, 0.0009, 0.0008, 0.0008, 0.0009, 0.0009, 0.0009, 0.0009, 0.0007, 0.0006, 0.0005, 0.0004, 0.0009, 0.0013, 0.0015, 0.0017, 0.0013, 0.0011, 0.0011)$.

Using (5), we get the following sum of details: $D^2_1 + D^2_2 = (0.0003, 0.0003, 0.0001, -0.0003, -0.0002, 0.0010, -0.0011, -0.0004, 0.0003, 0.0013, -0.0007, 0.0006, 0.0005, -0.0014, -0.0006, 0.0003, 0.0003, 0.0013, -0.0010, -0.0005, 0.0004)$.

The structure of this WRM gives a great opportunity to transit the extremum to another region. For example, if we want to eliminate the maximum in the $17^{th}$ signal's value, and put new extremums in the $1^{st}$ and the $13^{th}$ ones, we can take the following approximation coefficients: $\tilde{a}^2_2 = (0.0032, 0.0032, 0, 0.0032, 0)$. In general, we could take any other coefficients. The particular choice depends on the task to be solved, and the structure of WRM.

Using (6), we get the following approximation: $\tilde{A}^2_2 = (0.0020, 0.0017, 0.0015, 0.0016, 0.0016, 0.0008, 0.0003, -0.0000, -0.0004, 0.0006, 0.0013, 0.0016, 0.0020, 0.0009, 0.0002, -0.0000, -0.0004, 0.0006, 0.0013, 0.0016)$.

Then, we can calculate a new concentration difference signal and new concentration signals: $\tilde{\delta}^2 = (0.0023, 0.0020, 0.0017, 0.0013, 0.0014, 0.0018, -0.0008, -0.0005, -0.0002, 0.0019, 0.0006, 0.0022, 0.0025, -0.0005, -0.0004, 0.0003, 0.0008, -0.0003, 0.0008, 0.0020)$; $\tilde{c}^1_{(2)} = (0.0273, 0.0270, 0.0284, 0.0306, 0.0289, 0.0267, 0.0210, 0.0249, 0.0266, 0.0290, 0.0293, 0.0308, 0.0333, 0.0319, 0.0357, 0.0364, 0.0331, 0.0363, 0.0351, 0.0353)$; $\tilde{c}^2_{(2)} = (0.0251, 0.0250, 0.0267, 0.0293, 0.0274, 0.0249, 0.0218, 0.0253, 0.0268, 0.0271, 0.0286, 0.0287, 0.0308, 0.0324, 0.0361, 0.0361, 0.0323, 0.0366, 0.0343, 0.0334)$.

Using these ratios and the quantities from the $2^{nd}$ row of Table 2, we can obtain new quantity signals $\hat{q}^1_{(2)}$ and $\hat{q}^2_{(2)}$.

As we've done before, we need to multiply these quantity signals by the coefficients $\sum_{i=1}^{20} q^1_i / \sum_{i=1}^{20} \hat{q}^1_{(2)_i} = 0.9929$ and $\sum_{i=1}^{20} q^2_i / \sum_{i=1}^{20} \hat{q}^2_{(2)_i} = 0.9936$ to preserve signals' mean values. And the last thing to complete is to round the signal.

The results can be found in Table 2 (the last 4 rows).

Also, to compare the results obtained by using different wavelet bases, we presented the initial and two final concentration difference signals in Fig. 2.

It is important to note that rounding the quantities can lead to changes in wavelet decomposition details. But, in most cases these changes are not very significant and don't pose a big threat to the data utility preserving.

All that is left to complete the task is to construct a new microfile. We can always do that by changing vital values of different records in order to gain needed distribution.

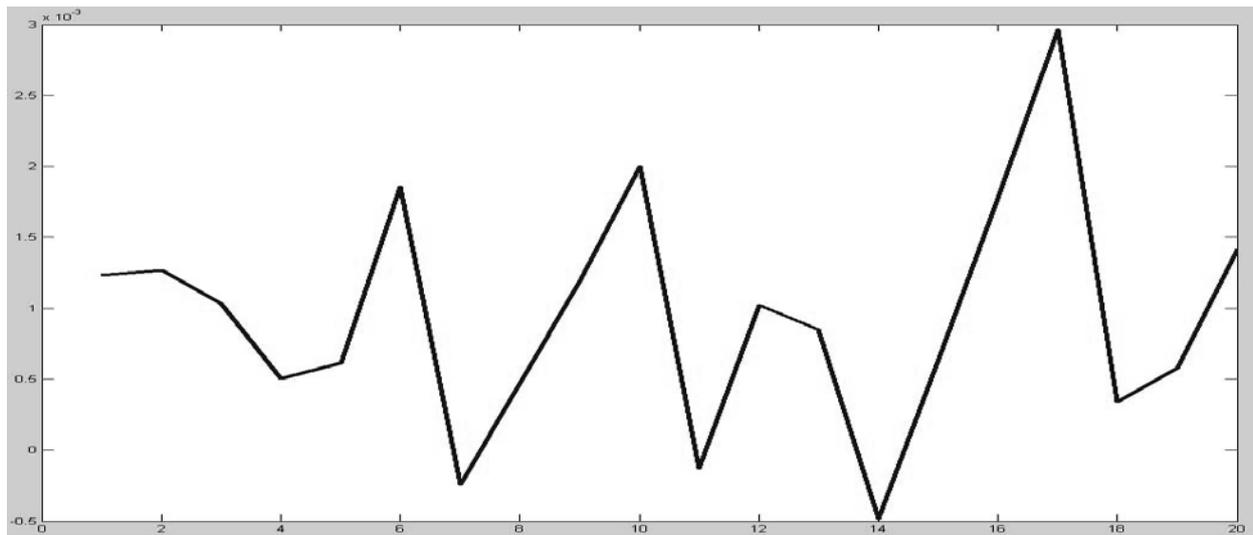

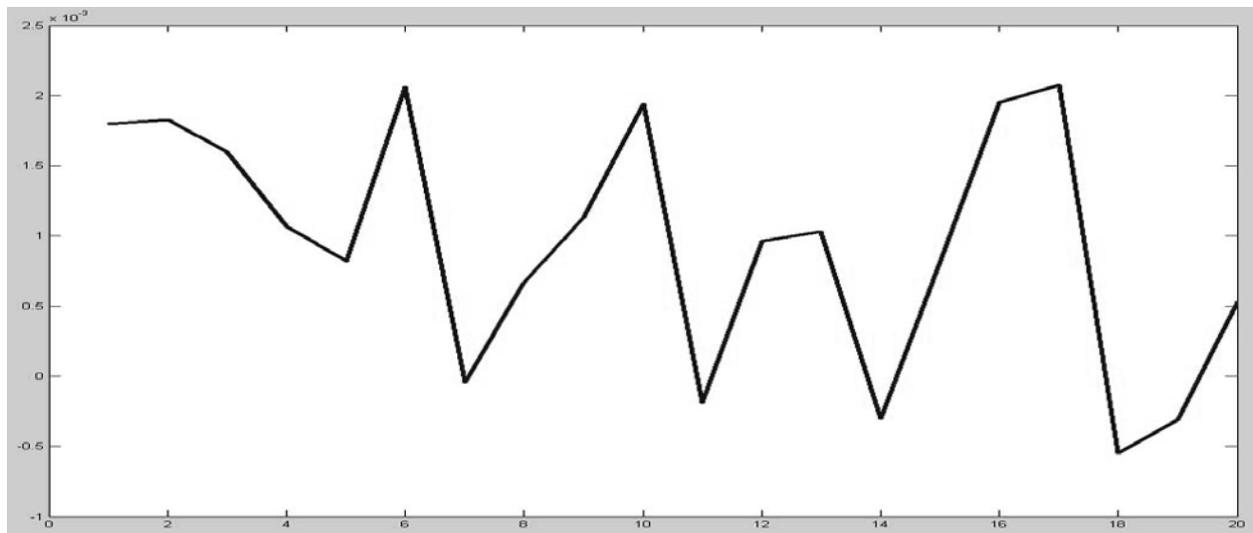

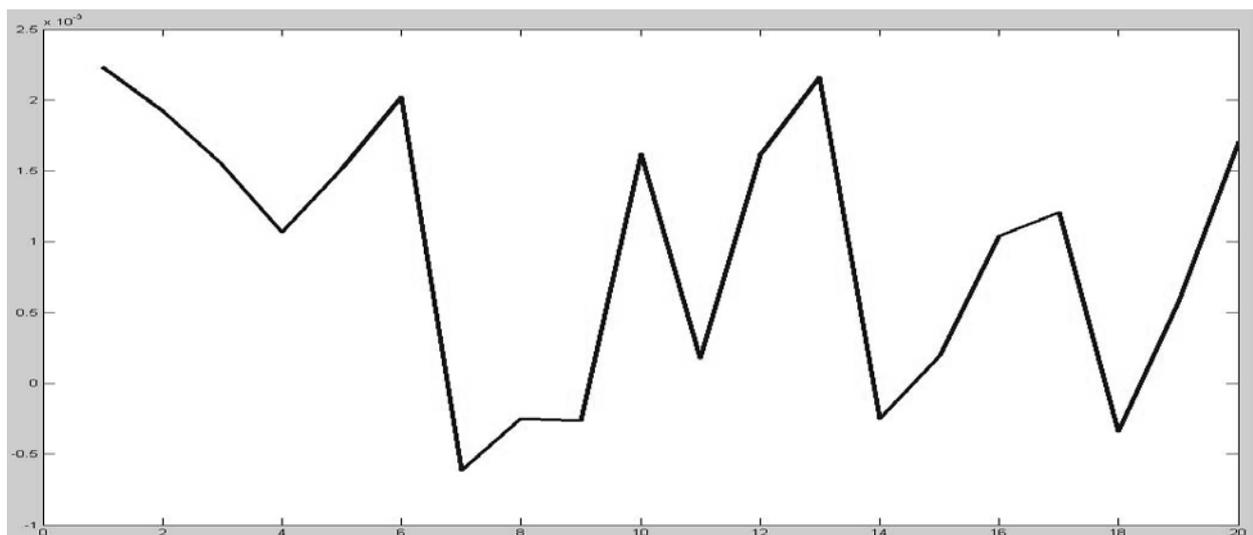

Fig. 2. Concentration difference signals: a) the initial one; b) the new signal obtained using the first order wavelet filter; c) the new signal obtained using the second order one.

Table 2. Quantities and ratios distributed by regions.

| Region code | 1P | 1V | 3 | 4 | 5 | 6 | 7 |
|---|---|---|---|---|---|---|---|
| All people | 220952 | 6326 | 474894 | 49411 | 238279 | 61883 | 82198 |
| Males (initial) | 5808 | 166 | 13164 | 1474 | 6683 | 1655 | 1727 |
| Signal $c^1$ | 0.0263 | 0.0262 | 0.0277 | 0.0298 | 0.0280 | 0.0267 | 0.0210 |
| Females (initial) | 5535 | 158 | 12671 | 1449 | 6536 | 1540 | 1747 |
| Signal $c^2$ | 0.0251 | 0.0250 | 0.0267 | 0.0293 | 0.0274 | 0.0249 | 0.0213 |
| Signal $\hat{c}^1$ (1) | 0.0269 | 0.0268 | 0.0283 | 0.0304 | 0.0282 | 0.0269 | 0.0210 |
| Males (final 1) | 5900 | 169 | 13359 | 1494 | 6694 | 1658 | 1718 |
| Signal $\hat{c}^2$ (1) | 0.0251 | 0.0250 | 0.0267 | 0.0293 | 0.0274 | 0.0249 | 0.0211 |
| Females (final 1) | 5516 | 157 | 12627 | 1444 | 6513 | 1535 | 1724 |
| Signal $\hat{c}^1$ (2) | 0.0273 | 0.0270 | 0.0284 | 0.0306 | 0.0289 | 0.0267 | 0.0210 |
| Males (final 2) | 5996 | 169 | 13368 | 1500 | 6826 | 1642 | 1715 |
| Signal $\hat{c}^2$ (2) | 0.0251 | 0.0250 | 0.0267 | 0.0293 | 0.0274 | 0.0249 | 0.0218 |
| Females (final 2) | 5500 | 157 | 12590 | 1440 | 6494 | 1530 | 1785 |

| Region code | 8 | 9 | 10 | 11 | 12 | 13 | 14 |
|---|---|---|---|---|---|---|---|
| All people | 208428 | 183928 | 43037 | 76918 | 268221 | 65895 | 16548 |
| Males (initial) | 5183 | 4890 | 1251 | 2191 | 7961 | 2084 | 528 |
| Signal $c^1$ | 0.0249 | 0.0266 | 0.0291 | 0.0285 | 0.0297 | 0.0316 | 0.0319 |
| Females (initial) | 5086 | 4671 | 1165 | 2201 | 7687 | 2028 | 536 |
| Signal $c^2$ | 0.0244 | 0.0254 | 0.0271 | 0.0286 | 0.0287 | 0.0308 | 0.0324 |
| Signal $\hat{c}^1$ (1) | 0.0251 | 0.0265 | 0.0290 | 0.0285 | 0.0296 | 0.0318 | 0.0319 |
| Males (final 1) | 5196 | 4853 | 1242 | 2179 | 7902 | 2084 | 525 |
| Signal $\hat{c}^2$ (1) | 0.0244 | 0.0254 | 0.0271 | 0.0287 | 0.0287 | 0.0308 | 0.0322 |
| Females (final 1) | 5068 | 4655 | 1161 | 2198 | 7660 | 2021 | 531 |
| Signal $\hat{c}^1$ (2) | 0.0249 | 0.0266 | 0.0290 | 0.0293 | 0.0308 | 0.0333 | 0.0319 |
| Males (final 2) | 5146 | 4855 | 1239 | 2234 | 8210 | 2177 | 524 |
| Signal $\hat{c}^2$ (2) | 0.0253 | 0.0268 | 0.0271 | 0.0286 | 0.0287 | 0.0308 | 0.0324 |
| Females (final 2) | 5249 | 4889 | 1158 | 2187 | 7638 | 2015 | 532 |

| Region code | 15 | 16 | 17 | 18 | 19 | 20 | Mean |
|---|---|---|---|---|---|---|---|
| All people | 299790 | 210976 | 31368 | 105710 | 260549 | 85428 | |
| Males (initial) | 11020 | 7990 | 1105 | 3832 | 9095 | 2971 | 4538.9 |
| Signal $c^1$ | 0.0368 | 0.0379 | 0.0352 | 0.0363 | 0.0349 | 0.0348 | 0.0302 |
| Females (initial) | 10827 | 7616 | 1012 | 3796 | 8945 | 2850 | 4402.8 |
| Signal $c^2$ | 0.0361 | 0.0361 | 0.0323 | 0.0359 | 0.0343 | 0.0334 | 0.0292 |
| Signal $\hat{c}^1$ (1) | 0.0369 | 0.0381 | 0.0343 | 0.0363 | 0.0349 | 0.0339 | 0.0303 |
| Males (final 1) | 11013 | 7984 | 1071 | 3811 | 9045 | 2879 | 4538.8 |
| Signal $\hat{c}^2$ (1) | 0.0361 | 0.0361 | 0.0323 | 0.0368 | 0.0352 | 0.0334 | 0.0293 |
| Females (final 1) | 10789 | 7589 | 1008 | 3876 | 9144 | 2840 | 4402.8 |
| Signal $\hat{c}^1$ (2) | 0.0357 | 0.0364 | 0.0331 | 0.0363 | 0.0351 | 0.0353 | 0.0304 |
| Males (final 2) | 10638 | 7618 | 1031 | 3805 | 9087 | 2997 | 4538.9 |
| Signal $\hat{c}^2$ (2) | 0.0361 | 0.0361 | 0.0323 | 0.0366 | 0.0343 | 0.0334 | 0.0294 |
| Females (final 2) | 10758 | 7567 | 1006 | 3843 | 8888 | 2832 | 4402.9 |

## 5. Conclusion and Future Research

In the paper, we discussed a completely new approach to providing data group anonymity which is most acceptable for hiding relative distributions and comparative patterns. The proposed method can be considered as a method complementary to the existing ones which in fact solve providing individual anonymity problem only. Also, we showed that some real-life tasks can be completed by redistributing appropriate ratio differences which is a totally novel approach to providing data anonymity.

Another conclusion is that different wavelet bases may yield completely different results. Besides, it was clearly viewed that some wavelet bases serve well when we need to transit the maximum signal values, and the others are most acceptable for creating alleged signal extremums.

Though, there are still other problems not solved yet. In our opinion, some of them doubtlessly are:
- The problem of choosing optimal wavelet base is still open.
- It is important to introduce group anonymity measure to be able to evaluate data utility loss.